\documentclass[12pt]{article}
\usepackage{amssymb,graphicx}
\newcommand{\adoa}[1]{\frac{\dot a_{#1}}{a_{#1}}}
\newcommand{\addoa}[1]{\frac{\ddot a_{#1}}{a_{#1}}}

\begin{document}

\begin{flushright}
{\tt KIAS-P05015}
\end{flushright}

\vspace{5mm}

\begin{center}
{{\Large \bf Cosmology of Antisymmetric Tensor Field\\[2mm]
in D-brane Universe}\\[12mm]
Eung Jin Chun\\
{\it Korea Institute for Advanced Study,\\
207-43 Cheongryangri 2-dong Dongdaemun-gu, Seoul 130-722, Korea}\\
{\tt ejchun@kias.re.kr}\\[4mm]
Hang Bae Kim\\
{\it Department of Physics,
Hanyang University, Seoul 133-791, Korea}\\
{\tt hbkim@hanyang.ac.kr}\\[4mm]
Yoonbai Kim\\
{\it BK21 Physics Research Division and Institute of Basic Science,\\
Sungkyunkwan University, Suwon 440-746, Korea}\\
{\tt yoonbai@skku.edu}
}
\end{center}

\vspace{5mm}

\begin{abstract}
We analyze homogeneous, anisotropic cosmology driven by a self-interacting
``massive'' antisymmetric tensor field $B_{\mu\nu}$ which is
present in string theories with  D-branes.  Time-dependent
magnetic $B$ field existing in the early universe can lead to the
Bianchi type I universe. Evolutions of such a tensor field are
solved exactly or numerically in the universe dominated by vacuum
energy, radiation, and $B$ field itself. The matter-like behavior
of the $B$ field (dubbed as ``$B$-matter'') ensures that the
anisotropy disappears at late time and thus becomes unobservable
in a reasonable cosmological scenario.  Such a feature should be
contrasted to the cosmology of the conventional massless
antisymmetric tensor field.
\end{abstract}


\newpage

\setcounter{equation}{0}

\section{Introduction}

The antisymmetric tensor field (Kalb-Ramond field) $B_{\mu\nu}$ is
an indispensable field in effective theories derived from the
string theory. It was considered as a massless degree of freedom
in the bulk spacetime and thus it is dual to a pseudo-scalar field
$h$: $H_{\mu\nu\rho}=
\partial_{[\mu} B_{\nu\rho]} = \epsilon_{\mu\nu\rho\sigma}
\partial^\sigma  h$ in four dimensions~\cite{Kalb:1974yc}.
Once we take into account D-brane which easily forms a bound state
of D$p$-brane-fundamental strings~\cite{Witten:1995im}, such a $B$
field alone becomes gauge non-invariant on the D$p$-brane and
gauge invariance is restored through its coupling to a U(1) vector
field $A_{\mu}$ living on a D$p$-brane, i.e., the gauge-invariant
field strength is
\begin{equation} \label{BplusF}
 B_{\mu\nu}+2\pi\alpha' F_{\mu\nu} \equiv {\cal B}_{\mu\nu},
 \quad\textrm{where}\quad F_{\mu\nu} = \partial_\mu A_\nu -\partial_\nu A_\mu .
\end{equation}
Note that ${\cal B}_{\mu\nu}$ is invariant under the tensor gauge
transformation
\begin{eqnarray}
\delta
B_{\mu\nu}=\partial_{\mu}\zeta_{\nu}-\partial_{\nu}\zeta_{\mu}\,,
\qquad
\delta A_{\mu}=-\zeta_{\mu}/2\pi\alpha' 
\nonumber
\end{eqnarray}
including the ordinary U(1) gauge transformation $\zeta_\mu =
\partial_\mu \lambda$.
Intriguingly enough, the effective action of antisymmetric tensor
field derived on the D-brane dictates this Abelian tensor field of
second rank to be massive and
self-interacting~\cite{Cremmer:1973mg,Witten:1995im,Pol}.

Another indispensable field is gravitational field $g_{\mu\nu}$
and, once it also enters into the playground in the string scale,
it is natural to deal with cosmological issues due to the
existence of the antisymmetric tensor field.
In this paper, we consider the
cosmology of the antisymmetric tensor field in a D-brane world
which is supposed to contain our universe. The cosmological
solutions with the massive antisymmetric tensor field ${\cal B}$
are quite different from those with the massless one. In the
latter case, a homogenous and isotropic universe can be realized
for $H_{0ij}=0$ and $H_{ijk} = f \epsilon_{ijk} $ with some
constant $f\propto
\partial_t h$ \cite{Goldwirth:1993ha}.
Another homogeneous solution leading to
an anisotropic Bianchi type I universe takes the configuration of
$H_{ijk}=0$ and $H_{0ij}=\epsilon_{ijk} L_k$ with one
non-vanishing constant $L_k \propto \partial_k h$
\cite{copeland:1995}.  Such an anisotropy could be in conflict
with observation.  Indeed, the anisotropy driven by a
non-vanishing $L_k$  grows in an expanding universe
\cite{copeland:1995}.
In relation with anisotropy in Bianchi type universes,
various issues have been discussed \cite{Barrow:2002}.
In the brane world where the antisymmetric
tensor field becomes massive, time-dependent homogeneous solutions
are described by ``magnetic'' degrees of freedom ${\cal B}_{ij} =
{\cal B}_{ij}(t)$ leading to $H_{ijk}=0$ and $H_{0ij}=\partial_t
{\cal B}_{ij}$, which does not allow isotropic cosmology. Assuming
the simplest form of  anisotropic universe, Bianchi type I,
we will analyze the cosmology with the massive magnetic  field
${\cal B}_{ij}$  in various situations of inflationary, radiation dominated,
or  ${\cal B}$-dominated era. As we will see, the initial
anisotropy is damped away and thus a practically isotropic
universe is recovered at late time due to the ordinary matter-like
behavior of the ${\cal B}$ field.

The paper is organized as follows.  In Section 2, we describe the
four-dimensional low-energy effective action of D-brane world from
which the mass of the antisymmetric tensor field  and its
equations of motion coupled to the gravity are derived. In Section
3, we find time-dependent homogenous solutions with one magnetic
component ${\cal B}_{12}(t) \neq 0$ which leads to the Bianchi
type I cosmology.  Section 4 concludes the paper.

\section{\large\bf Antisymmetric Tensor Field on D-brane}

Suppose that our universe is a part of a D$p$-brane. Then its
bosonic sector comprises bulk degrees including the graviton
$g_{\mu\nu}$, the dilaton $\Phi$, and the antisymmetric tensor field
$B_{\mu\nu}$ of rank-two in addition to U(1) gauge field $A_{\mu}$
living on the D$p$-brane. Combining $B_{\mu\nu}$ with $F_{\mu\nu}$
as in (\ref{BplusF}), the bosonic action of our system in string
frame is given by the sum of the bulk terms in ten dimensions;
\begin{eqnarray} \label{act10}
S_{10} = {1\over 2 \kappa_{10}^2} \int d^{10}x \, \sqrt{-g}\,
e^{-2\Phi} \left[ R -2\Lambda_{10}+ 4 (\partial \Phi)^{2} -{1\over
12} H^{2} \right],
\end{eqnarray}
and the brane terms in $p$ spatial dimensions;
\begin{eqnarray} \label{dpac}
S_{{\rm D}p} = -\mu_p \int d^{p+1} x\, e^{-\Phi}
\sqrt{-\det(g+{\cal B}) }
\, ,
\end{eqnarray}
where $\kappa_{10}^2={1\over2} (2\pi)^7 \alpha'^4$ and $\mu_p^2=
(\pi/\kappa_{10}^2) (4\pi^2\alpha')^{3-p}$ \cite{Pol}. Recall that
$\alpha'$ defines the string scale; $m_{{\rm s}} =
\alpha'^{-1/2}$. Here the spacetime indices  are not explicitly
expressed and the field strength $H$ of the antisymmetric tensor
field is again $H_{\mu\nu\rho}=\partial_{[\mu}{\cal B}_{\nu\rho]}$
because of Bianchi identity of the U(1) gauge field.

To consider the cosmology of our universe, let us compactify six
extra-dimensions with a common radius $R_{\rm c}$ and assume the
stabilized dilaton leading to a finite string coupling $g_{\rm
s}=e^\Phi$. Then, the four-dimensional effective action becomes
\begin{eqnarray}
S_{4}&=& {1\over 2\kappa_4^2} \int d^4x \sqrt{-g}
\left[ R-2\Lambda
-\frac{1}{12} H_{\mu\nu\rho}H^{\mu\nu\rho}\right. \nonumber\\
&& \hspace{15mm} \left. -{m_B^2} \sqrt{1 + \frac{1}{2} {\cal
B}_{\mu\nu} {\cal B}^{\mu\nu} - \frac{1}{16} \left( {\cal
B}^*_{\mu\nu} {\cal B}^{\mu\nu}\right)^2 } \right], \label{act1a}
\end{eqnarray}
where ${\cal B}^{\ast}_{\mu\nu}=
\sqrt{-g}\epsilon_{\mu\nu\alpha\beta} {\cal B}^{\alpha\beta}/2$
with $\epsilon_{0123} = 1$. Here the reduced four-dimensional
Planck length $\kappa_4$ is determined by $1/\kappa_4^2=R_{{\rm
c}}^6/g_{{\rm s}}^2 \kappa_{10}^2$.  The effective action
(\ref{act1a}) up to the quadratic terms shows that the
antisymmetric tensor field acquires the mass;
\begin{equation}\label{massb}
m_B=  \pi^{1\over4} \alpha_{{\rm s}}^{p-3\over16}
\left( m_{{\rm c}} \over m_{{\rm P}} \right)^{{15-p \over8}} m_{{\rm P}},
\end{equation}
where $\alpha_{\rm s} = g^2_{\rm s}/4\pi$, $m_{{\rm c}}=1/R_{{\rm
c}}$, and $m_{{\rm P}}=1/\kappa_4=2.4\times10^{18}$  GeV. As we
will see, the massive ${\cal B}$ field  (named as $B$-matter)
behaves like a massive scalar in some respects in its cosmological
evolution, due to which the intrinsic anisotropy set by
directional ${\cal B}$ field gets diminished eventually.


From the action (\ref{act1a}),
one finds the following equations of motion for the metric $g_{\mu\nu}$
and the antisymmetric tensor field ${\cal B}_{\mu\nu}$
\begin{equation}
\label{B-eq}
\nabla^\mu H_{\mu\nu\rho}
-  m_B^2
\frac{{\cal B}_{\nu\rho}-\frac{1}{4}{\cal B}^*_{\nu\rho} ({\cal B}^*{\cal B})}
{\sqrt{1+\frac12{\cal B}^2-\frac{1}{16}({\cal B}^*{\cal B})^2}}
= 0,
\end{equation}
\begin{equation}
\label{A-eq}
\nabla^\mu\left[
\frac{{\cal B}_{\mu\nu}-\frac14{\cal B}^*_{\mu\nu}({\cal B}^*{\cal B})}{
\sqrt{1+\frac12{\cal B}^2-\frac{1}{16}({\cal B}^*{\cal B})^2}}\right]=0,
\end{equation}
\begin{equation}
\label{E-eq}
G_{\mu\nu} = \Lambda g_{\mu\nu} +
\frac{1}{12}\left(3H_{\mu\lambda\rho}H_\nu^{\phantom{\nu}\lambda\rho}
-\frac12g_{\mu\nu}H^2\right)
+\frac{m_B^2}{2}
\frac{{\cal B}_{\mu\lambda}{\cal B}_\nu^\lambda
-g_{\mu\nu}(1+\frac12{\cal B}^2)}
{\sqrt{1+\frac12{\cal B}^2-\frac{1}{16}({\cal B}^*{\cal B})^2}},
\end{equation}
where $\nabla_\mu$ is the covariant derivative. The equation
(\ref{A-eq}) is also required from the consistency of
(\ref{B-eq}), as you can easily check by applying $\nabla^\nu$ to
(\ref{B-eq}). Out of six equations in (\ref{B-eq}), one can find
that three of them are dynamical equations and the other three are
constraints. Therefore, the field ${\cal B}_{\mu\nu}$ contains
only three degrees of freedom as anticipated from the discussion
of (\ref{BplusF}).

\setcounter{equation}{0}
\section{Time-dependent Homogeneous Solutions}
In this section we investigate evolutions of homogeneous magnetic
components of the antisymmetric tensor field ${\cal B}_{ij}$ with
and without gravity. In the flat spacetime composed of a D-brane we
find an oscillating solution. In the B-dominated D-brane universe,
isotropy is recovered despite of the initially-assumed anisotropy,
whereas it is not without D-brane.

\subsection{Flat spacetime: Homogeneous magnetic $B$-matter}
We begin this subsection by turning off the gravitational coupling
on the D-brane.
We now consider homogeneous configuration ${\cal B}_{\mu\nu}(t)$
and look for its time evolution in flat spacetime with
$g_{\mu\nu}=\eta_{\mu\nu}$. Time component of the equation
(\ref{A-eq}) is trivially satisfied and spatial components become
\begin{equation}\label{hfA}
\partial_{0}\left[
\frac{{\bf E}-{\bf B}({\bf E}\cdot{\bf B})}{\sqrt{1-{\bf E}^{2}+{\bf B}^{2}
-({\bf E}\cdot {\bf B})^{2}}} \right]=0,
\end{equation}
where ${\cal B}_{i0}\equiv ({\bf E})^{i}$ and ${\cal B}_{ij}\equiv
\epsilon_{0ijk}({\bf B})^{k}$. This equation (\ref{hfA}) combined
with the $0i$-component of (\ref{B-eq}) reads
\begin{equation}
 {\bf E} = {\bf B} ({\bf E}\cdot{\bf B}) \,.
\end{equation}
This allows a non-trivial solution with ${\bf E}=0$ and ${\bf
B}(t)\neq0$. Then, the dynamical equation of the ${\bf B}$ field
is summarized by
\begin{equation}\label{seqB}
\partial_{0}^{2}{\bf B}= -m_{B}^{2}
\frac{{\bf B}}{\sqrt{1+{\bf B}^{2}}}.
\end{equation}
Integration of the equation (\ref{seqB}) gives
\begin{equation}\label{foeq}
{\cal E}=\frac{1}{2m_{B}^{2}}{\dot {\bf B}}^{2}+V_{{\rm eff}}({\bf
B}),
\end{equation}
where ${\cal E}$ is an integration constant, $V_{{\rm eff}}({\bf
B})=\sqrt{1+{\bf B}^{2}}\,$, and overdot is differentiation over
time variable $t$. Let us assume that the direction of the magnetic field
is fixed, i.e., ${\hat {\bf k}}$ defined by ${\bf B}(t)=B(t){\hat {\bf
k}}$ is independent of time. Then (\ref{foeq}) is rewritten by an
integral equation
\begin{equation}\label{Bs2}
m_{B}(t-t_{0})=\pm\int_{B_{0}}^{B}
\frac{dB}{\sqrt{2({\cal E}-\sqrt{1+B^{2}}\, )}}.
\end{equation}
The only pattern of nontrivial solutions is oscillating one with
amplitude $\sqrt{{\cal E}^{2}-1}$ for ${\cal E}>1$. This
oscillating solution with fixed amplitude is natural at classical
level in flat spacetime when a homogeneous condensation of
magnetic field is given by an initial condition. In expanding
universes, the solution is expected to change to that with
oscillation and damping due to growing of the spatial scale
factor, which means the dilution of $B$-matter.

\subsection{B-dominated universe}

We are interested in the cosmological implications of the large
scale fluxes of the antisymmetric tensor field, which might have
existed in the early universe. Thus we look for cosmological
solutions of (\ref{B-eq})--(\ref{E-eq}) with appropriate initial
conditions and examine the effects on the cosmological evolution.
First we consider the case that the universe is dominated by the
massive tensor field. Extending the above flat spacetime solution,
we take into account the homogeneous configurations with an ansatz
${\cal B}_{ij}={\cal B}_{ij}(t)$ and  ${\cal B}_{0i}=0$. For
general non-vanishing ${\cal B}_{ij}(t)$, the right-hand side of
(\ref{E-eq}) says that the energy-momentum tensor $T_{\mu\nu}$ of
antisymmetric tensor field will have non-vanishing off-diagonal
components $T_{ij}$. To consider the simplest form of anisotropic
cosmology,
we take only single component of ${\cal B}_{ij}$ to be nonzero;
namely ${\cal B}_{12}(t)\equiv B(t)$ and ${\cal B}_{23}(t)={\cal
B}_{31}(t)=0$. Even in this case, $T_{11}(t)=T_{22}(t)\ne
T_{33}(t)$ in general and thereby the corresponding metric is not
isotropic. Therefore, we take the metric to be of Bianchi type I
\begin{equation}
ds^2 = -dt^2 + a_1(t)^2(dx^1)^2 + a_2(t)^2(dx^2)^2 + a_3(t)^2(dx^3)^2.
\end{equation}
The energy-momentum tensor of the $B$ field is written in the
form
\begin{eqnarray} \label{TB}
{(T_B)_\mu}^\nu &=& {(T_{\rm bulk})_\mu}^\nu + {(T_{\rm brane})_\mu}^\nu \\
&=& \kappa_4^{-2}\Lambda{\delta_\mu}^\nu + {\rm
diag}\left[-\rho_B,-\rho_B,-\rho_B,+\rho_B\right] + {\rm
diag}\left[-\rho_b,-\tilde\rho_b,-\tilde\rho_b,-\rho_b\right],
\nonumber
\end{eqnarray}
where $\rho_B=\dot B^2/4\kappa_4^2a_1^2a_2^2$,
$\rho_b=(m_B^2/2\kappa_4^2)\sqrt{1+B^2/a_1^2a_2^2} \,$ and
$\tilde\rho_b=(m_B^2/2\kappa_4^2)/\sqrt{1+B^2/a_1^2a_2^2}\,$.
Thus, the Einstein equations (\ref{E-eq}) for the
tensor-field-dominated case are
\begin{eqnarray}
\adoa1\adoa2+\adoa2\adoa3+\adoa3\adoa1 &=& \kappa_4^2\rho_B+\kappa_4^2\rho_b+\Lambda,
\label{ei1}\\
\addoa2+\addoa3+\adoa2\adoa3 &=& \kappa_4^2\rho_B+\kappa_4^2\tilde\rho_b+\Lambda, \\
\addoa3+\addoa1+\adoa3\adoa1 &=& \kappa_4^2\rho_B+\kappa_4^2\tilde\rho_b+\Lambda, \\
\addoa1+\addoa2+\adoa1\adoa2 &=&
-\kappa_4^2\rho_B+\kappa_4^2\rho_b+\Lambda. \label{ei4}
\end{eqnarray}
The equation of motion for $B(t)$ is
\begin{equation}
\label{Beq} \ddot B + \left(-\frac{\dot a_1}{a_1}-\frac{\dot
a_2}{a_2}+\frac{\dot a_3}{a_3}\right)\dot B
+m_B^2\frac{B}{\sqrt{1+B^2/a_1^2a_2^2}}=0.
\end{equation}

Introducing $\alpha_i=\ln a_i$, we rewrite the Einstein equations
(\ref{ei1})--(\ref{ei4}) as
\begin{eqnarray}
\dot\alpha_1\dot\alpha_2+\dot\alpha_2\dot\alpha_3+\dot\alpha_3\dot\alpha_1
&=& \kappa_4^2\rho_B+\kappa_4^2\rho_b+\Lambda,
\label{00}\\
\ddot\alpha_1+\dot\alpha_1(\dot\alpha_1+\dot\alpha_2+\dot\alpha_3)
&=& \kappa_4^2\rho_b+\Lambda,
\label{11}\\
\ddot\alpha_2+\dot\alpha_2(\dot\alpha_1+\dot\alpha_2+\dot\alpha_3)
&=& \kappa_4^2\rho_b+\Lambda,
\label{22}\\
\ddot\alpha_3+\dot\alpha_3(\dot\alpha_1+\dot\alpha_2+\dot\alpha_3)
&=& 2\kappa_4^2\rho_B+\kappa_4^2\tilde\rho_b +\Lambda.
\label{33}
\end{eqnarray}
Subtraction of (\ref{22}) from (\ref{11}) gives
\begin{eqnarray}\label{12}
\ddot\alpha_1 - \ddot\alpha_2 +(\dot\alpha_1 - \dot\alpha_2 )
(\dot\alpha_1 + \dot\alpha_2 + \dot\alpha_3 )=0.
\end{eqnarray}
A natural solution is $\dot\alpha_1=\dot\alpha_2$
which, with the help of scaling of $x^{1}$ and $x^{2}$ coordinates,
leads to the isotropy in $x^{1}x^{2}$-plane,
$\alpha_1=\alpha_2$ or equivalently $a_1=a_2$.

\subsubsection{Massless limit ($m_B=0$)}

To see the effect of the brane on the spacetime dynamics, let us
first consider the limit of $m_B=0$ which corresponds to either the
absence of the brane or the limit of vanishing string coupling
$g_{{\rm s}}$, namely the usual massless antisymmetric
tensor field. In this limit, the equation (\ref{Beq}) is easily
integrated to yield a constant of motion
\begin{equation} \label{L3}
\frac{a_3\dot B}{a_1a_2} \equiv L_3 \ (\textrm{constant}).
\end{equation}
With the vanishing potential, $B$ manifests itself by
non-vanishing time derivatives. In the dual variable, it
corresponds to the homogeneous gradient along $x^3$-direction. The
spacetime evolution with the dilaton rolling in this case was
studied in Ref.~\cite{copeland:1995}. Here we have assumed that
the dilaton is stabilized by some mechanism.  Following
Ref.~\cite{copeland:1995} we introduce a new time coordinate
$\lambda$ via the relation $d\lambda=L_3dt/a_1a_2a_3$. Then the
equations (\ref{00})--(\ref{33}) can be written as
\begin{eqnarray}
\label{Eeq123}
\alpha_1'\alpha_2'+\alpha_2'\alpha_3'+\alpha_3'\alpha_1'
&=& \frac14a_1^2a_2^2, \\
\alpha_1''=\alpha_2'' &=& 0, \\
\label{Eeq333}
\alpha_3'' &=& \frac12a_1^2a_2^2,
\end{eqnarray}
where the prime denotes the differentiation with respect to $\lambda$.

The solutions for $\alpha_1$ and $\alpha_2$ are trivial
\begin{equation}
\alpha_1 = C_1\lambda, \quad
\alpha_2 = C_2\lambda,
\end{equation}
where $C_{1,2}$ are constants and we omitted the integration
constants corresponding simply to re-scaling of scale factors. The
$\alpha_3$-equation (\ref{Eeq333}) is also easily integrated to
give
\begin{equation}
\alpha_3 = \frac{e^{2(C_1+C_2)\lambda}}{8(C_1+C_2)^2}+C_3\lambda .
\end{equation}
The constraint equation (\ref{Eeq123}) relates $C_{1,2}$ and $C_3$ by
$C_3=-C_1C_2/(C_1+C_2)$.
Then the relation between $\lambda$ and $L_3t$ is explicitly given by
\begin{eqnarray}
L_3t &=& \int^\lambda d\lambda\; a_1(\lambda)a_2(\lambda)a_3(\lambda) \nonumber\\
&=& \int^\lambda d\lambda \exp\left[
(C_1+C_2+C_3)\lambda+\frac{e^{2(C_1+C_2)\lambda}}{8(C_1+C_2)^2}\right]
\nonumber\\
&=& \left[8(C_1+C_2)\right]^{\frac{C_1+C_2+C_3}{C_1+C_2}}
\int^x dy\; y^{-P}e^y,
\end{eqnarray}
where $x=e^{2(C_1+C_2)\lambda}/8(C_1+C_2)^2$ and $P=(C_1+C_2-C_3)/2(C_1+C_2)$.
The evolution of scale factors for large $L_3t$ is given by
\begin{equation}
a_{1,2} \propto \left(\log L_3t\right)^{q_{1,2}}, \qquad
a_3 \propto L_3t,
\end{equation}
where $q_{1,2}=C_{1,2}/2(C_1+C_2)$. Therefore, with non-vanishing
$B_{12}(t)$, only $a_3$ grows significantly and the spatial
anisotropy develops.  This can be seen clearly by considering the
ratio $H_{3}/H_{1,2}$ where $H_i\equiv \dot{a}_i/a_i$.  Taking
$C_1=C_2$ and thus $q_{1,2}=1/4$, we get
\begin{equation}
 {H_{3}\over H_{1,2}}= {4 \log L_3 t}
\end{equation}
which grows as time elapses. Finally, we remark that such
anisotropy cannot be overcome by some other type of isotropic
energy density in an expanding universe.  Assume the isotropic
universe ($a_i=a$) driven by, e.g, radiation energy density
$\rho_{{\rm R}}$.  Then, one finds $\rho_B \propto 1/a^2$ from
(\ref{TB}) and (\ref{L3}) and thus $\rho_B/\rho_{{\rm R}} \propto
a^2$, which implies that the late-time isotropic solution can be
realized only in a contracting universe \cite{copeland:1995}.

\subsubsection{The $B$ oscillation}

Let us now take into consideration the effect of space-filling
D-brane. To get sensible solution, we fine-tune the bulk
cosmological constant term to cancel the brane tension, that is
$\Lambda=-m_B^2/2$, so that the effective four-dimensional
cosmological constant vanishes. It is convenient to define a new
variable $b\equiv B/a_1a_2$. Then we can rewrite the full
equations as follows
\begin{eqnarray}
\label{Eeq-11}
\dot\alpha_1^2+2\dot\alpha_1\dot\alpha_3 &=& \frac14\left(\dot b+2\dot\alpha_1b\right)^2
+\frac{m_B^2}{2}\left(\sqrt{1+b^2}-1\right), \\
\label{Eeq-1}
\ddot\alpha_1+\dot\alpha_1\left(2\dot\alpha_1+\dot\alpha_3\right) &=&
\frac{m_B^2}{2}\left(\sqrt{1+b^2}-1\right), \\
\label{Eeq-3}
\ddot\alpha_3+\dot\alpha_3\left(2\dot\alpha_1+\dot\alpha_3\right)
&=& \frac12\left(\dot b+2\dot\alpha_1b\right)^2+
\frac{m_B^2}{2}\left(\frac{1}{\sqrt{1+b^2}}-1\right),
\end{eqnarray}
\begin{equation}
\label{Beq-2}
\ddot b+\left(2\dot\alpha_1+\dot\alpha_3\right)\dot b+
\left(2\ddot\alpha_1+2\dot\alpha_1\dot\alpha_3+\frac{m_B^2}{\sqrt{1+b^2}}\right)b=0.
\end{equation}
With the help of (\ref{Eeq-11}), (\ref{Beq-2}) can be written as
\begin{equation}
\label{Beq-3}
\ddot b+\left(2\dot\alpha_1+\dot\alpha_3\right)\dot b+
\left[-4\dot\alpha_1^2+m_B^2\left(\sqrt{1+b^2}-1+\frac{1}{\sqrt{1+b^2}}\right)\right]b=0,
\end{equation}
which is valid only for the $B$-dominated case, but more
convenient for numerical analysis. Here we have put
$\kappa_4\equiv1$.

First we examine the evolution of $b(t)$ and scale factors
qualitatively. {}From (\ref{Eeq-11})--(\ref{Beq-2}), we see that
the natural time scale is $m_B^{-1}$ and $m_Bt$ is the
dimensionless time variable. Suppose $b$ starts to roll from an
initial value $b_0$, while the universe is isotropic in the sense
that $\dot\alpha_{10}=\dot\alpha_{30}$. We assume initially
$a_{10}=a_{30}=1$ ($\alpha_{10}=\alpha_{30}=0$) and $\dot B_0=0$
so that $b_0=B_0$ and $\dot b_0=-2\dot\alpha_{10}B_0$. While $b$
is much larger than unity, the rapid expansion of $\alpha_1$ due
to the large potential proportional to $m_B^2b$ drives $b$ in
feedback to drop very quickly to a small value of order one. Our
numerical analysis in Figure~1 shows that this happens within
$m_Bt<2$ up to reasonably large value of $b_0$ for which the
numerical solution is working. The behavior of $b(t)$ after this
point is almost universal irrespective of the initial value $b_0$
if it is much larger than unity.

Once $b$ becomes smaller than unity, the quadratic term of mass
dominates over the expansion and $b$ begins to oscillate about
$b=0$. Then the expansion of the universe provides the slow
decrease of the oscillation amplitude. The situation is the same
as that of the coherently oscillating scalar field such as the
axion or the moduli in the expanding universe. For small $b$, the
energy-momentum tensor of the oscillating $B$ field is given by
${T_\mu}^\nu={\rm diag}[-\rho,p_1,p_2,p_3]$ where
\begin{eqnarray}
\rho &=& \frac14\left(\dot b+2 \dot\alpha_1b\right)^2
    + \frac12m_B^2\left(\sqrt{1+b^2}-1\right)
    \approx \frac14\left(\dot b^2+m_B^2b^2\right), \\
p_1=p_2 &=& -\frac14\left(\dot b+2\dot\alpha_1b\right)^2
    - \frac12m_B^2\left(\frac{1}{\sqrt{1+b^2}}-1\right)
    \approx -\frac14\left(\dot b^2-m_B^2b^2\right), \\
p_3 &=& \frac14\left(\dot b+2\dot\alpha_1b\right)^2
    - \frac12m_B^2\left(\sqrt{1+b^2}-1\right)
    \approx \frac14\left(\dot b^2-m_B^2b^2\right).
\end{eqnarray}
With the expansion of the universe neglected, the equation of
motion for $b$ is then approximated by
\begin{equation}
\label{Beq-osc}
\ddot b+m_B^2b\approx0.
\end{equation}
Since the oscillation is much faster than the expansion, we can
use the time-averaged quantities over one period of oscillation
for the evolution of spacetime. The equation (\ref{Beq-osc}) gives
the relation $\langle\dot b^2\rangle=\langle m_B^2b^2\rangle$.
Thus, the oscillating $B$ field has the property
$p_1,p_2,p_3\approx0$ and behaves like homogeneous and isotropic
matter. This justifies the name of {\em B-matter}. Therefore,
after $b$ begins to oscillate, the isotropy of the universe is
recovered.

To quantify how the universe recovers isotropy, let us define the
parameter
$$ s \equiv \sqrt{2} \frac{H_1-H_3}{2 H_1+H_3} .$$
When $H_1\approx H_3$, the evolution of the quantity $s(t)$ is
determined by
\begin{equation}\label{ani}
\dot{s} = \frac16\left( \frac{4p_1-p_3-9 H^2}{H}\right) s +
\frac13\frac{p_1-p_3}{H} + {\cal O}(s^2),
\end{equation}
where $H\equiv \sum_i H_i/3$. At late time, the equation
(\ref{ani}) is approximated as $\dot{s} \approx -(3/2) H s$ and
thus one finds $ s\propto 1/t$.  Since $H$ is also proportional to
$1/t$, the initial and final anisotropy $s_{i,f}$ follow the
relation
\begin{equation}
s_f = s_i \frac{H_f}{H_i},
\end{equation}
where $H_{i,f}$ denotes the initial and final Hubble parameter,
respectively. To get an idea of how fast the anisotropy
disappears, let us consider $H_f \sim 10^{-15}$ GeV which
corresponds to the Hubble parameter at the electroweak symmetry
breaking scale of temperature, $T\sim 100$ GeV. Taking the initial
condition $s_i\sim 1$ around the beginning of the $B$ oscillation
$H_i \sim m_B$, we find that the final anisotropy can be
completely neglected for reasonable value of $m_B$.

For $m_B t\gg1$ with $H_1\simeq H_2 \simeq H_3$ and $b\ll1$, the
asymptotic solution of (\ref{Eeq-11})--(\ref{Beq-2}) can be
explicitly found to yield
\begin{equation}
H\propto 2/3t\,, \quad b \propto 1/t\,,\quad \rho_{B,b} \propto
1/t^2
\end{equation}
which shows the usual matter-like evolution. That is, the total
energy density of the $B$-matter diminishes like $\rho \propto
1/a^3$ even though the amplitude $B(t)$ itself grows like
$B(t)\propto a^{1/2}$.

Our qualitative results can be confirmed by  solving the equations
(\ref{Eeq-1}), (\ref{Eeq-3}), and (\ref{Beq-3}) numerically.
Figure~1 shows the numerical solutions for the initial value
$b_0=100$. For the large value of $b_0$, $a_1$ grows very fast
while $a_3$ is frozen until $b$ becomes smaller than unity. Then
$b$ begins to oscillate and the universe becomes isotropic again
in that the expansion rates, $\dot\alpha_1$ and $\dot\alpha_3$,
converge and finally the universe becomes $B$-matter dominated.

\begin{figure}
\begin{center}
\includegraphics[height=180pt]{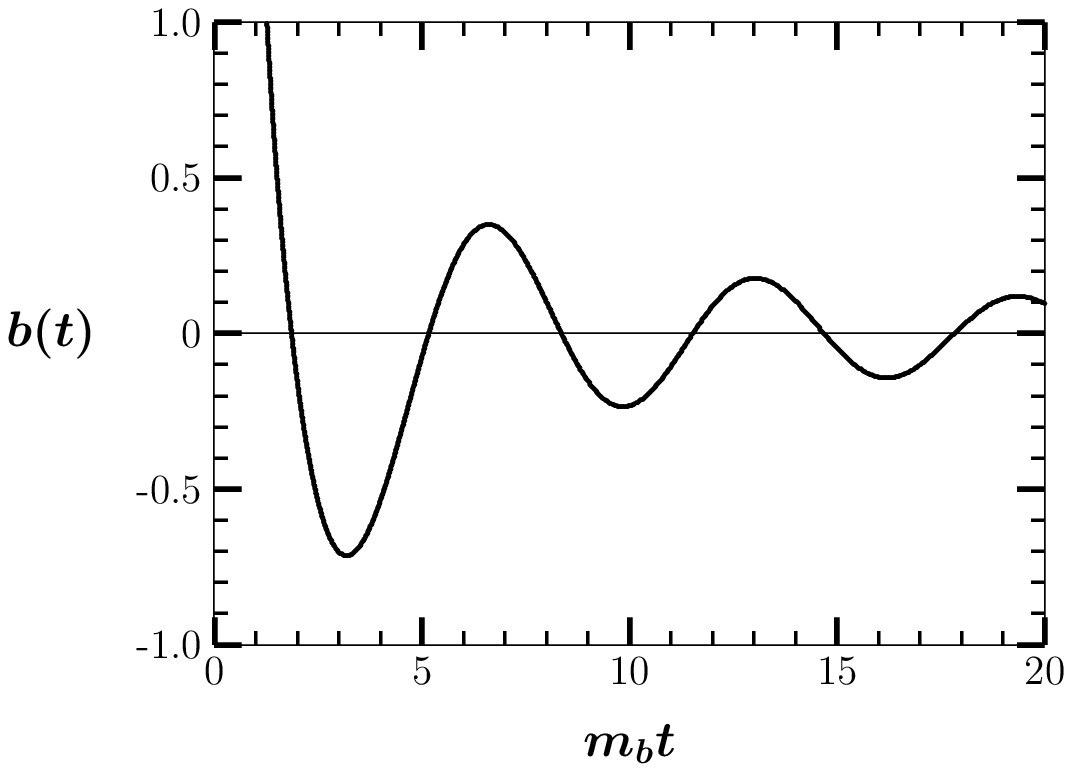}\\
\includegraphics[height=180pt]{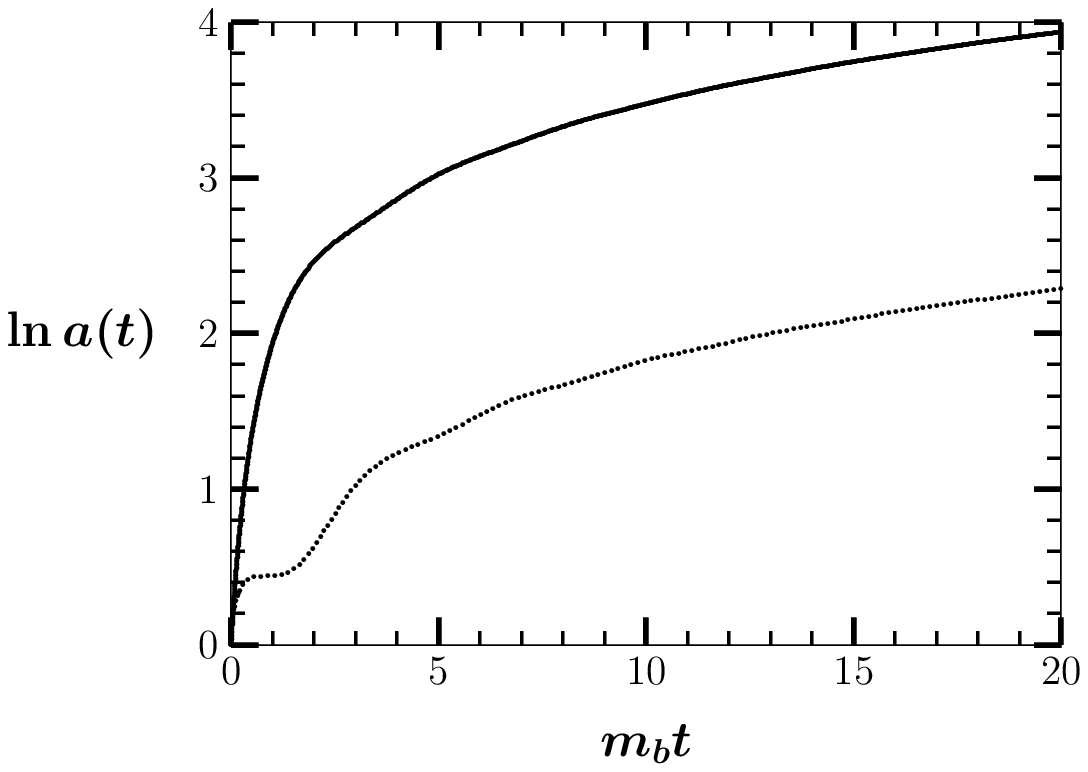}\\
\includegraphics[height=180pt]{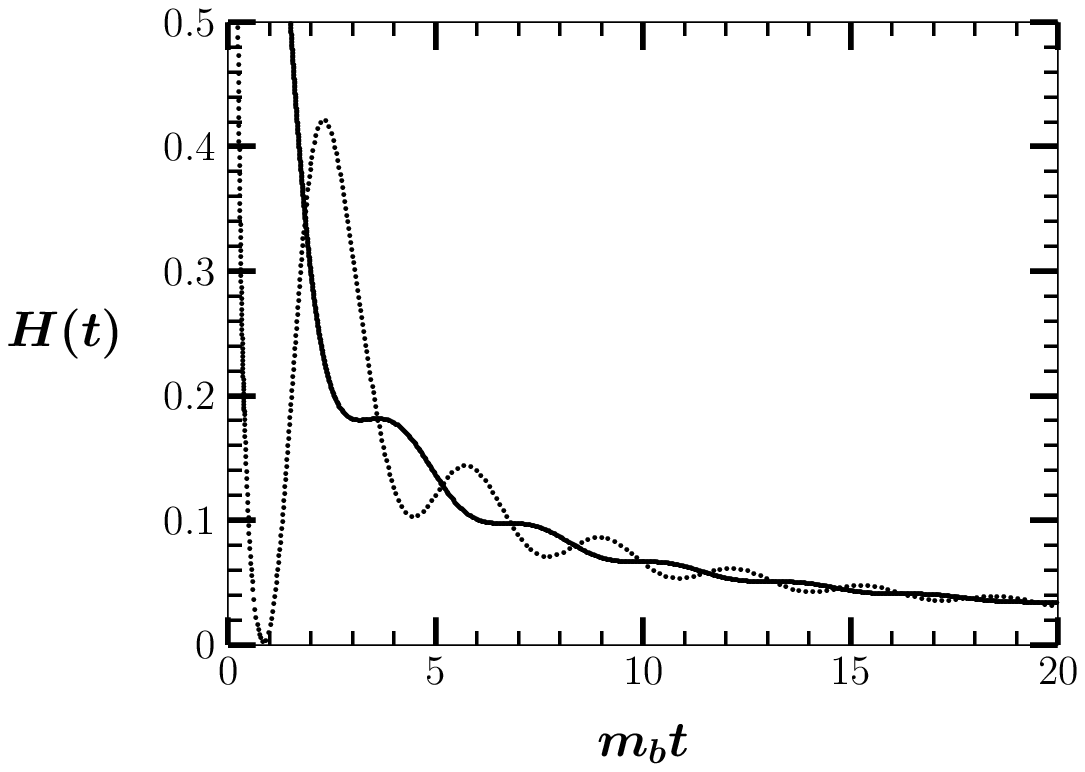}
\end{center}
\caption{Numerical solutions for $b_0=100$. (a) The evolution of $b(t)$.
(b) Scale factors $a_1(t)$ (solid line) and $a_3(t)$ (dotted line).
(c) Expansion rates $H_1(t)$ (solid line) and $H_3(t)$ (dotted line).}
\end{figure}

\subsection{Evolution of $B(t)$ in the RW background}

When the universe is dominated by homogeneous and isotropic matter
or radiation, while the $B$ field contributes a minor fraction,
the background geometry is described by the Robertson-Walker
metric and the equation for $b(t)$ is
\begin{equation}
\label{Beq-RW}
\ddot b + 3\frac{\dot a}{a}\dot b +
\left(2\frac{\ddot a}{a}+\frac{m_B^2}{\sqrt{1+b^2}}\right)b = 0.
\end{equation}
In this subsection, we study the evolution of $b(t)$ during inflation
and the radiation dominated era.

\subsubsection{Evolution during inflation}

During inflation, $\dot a/a\approx H$ (constant) and $\ddot
a/a\approx H^2$. We are assuming that the inflaton dominates over
the antisymmetric tensor field, which means $H^2\gg
m_B^2(\sqrt{1+b^2}-1)$.
Defining a dimensionless time variable as $Ht$, we write the
equation (\ref{Beq-RW}) as
\begin{equation}
\label{Beq-Inf1}
\ddot b+3\dot b+\left(2+\frac{m_B^2/H^2}{\sqrt{1+b^2}}\right)b=0.
\end{equation}
Due to $2H^2$ term, dominating over the oscillatory $m_B^2$ term,
$b$ exponentially dies away. Neglecting $m_B^2$ term, we have the
solution $b\approx b_0e^{-2Ht}$, corresponding to a frozen $B$
field at some value $B_0$, while its energy density decreases
exponentially during inflation. Therefore inflation dilutes away
the pre-existing $B$ field.

\subsubsection{Evolution during radiation dominated era}

During the radiation dominated era, $\dot a/a=1/2t$ and $\ddot
a/a=-1/4t^2$.
Defining the dimensionless time variable as $m_Bt$, we rewrite the
equation (\ref{Beq-RW}) as
\begin{equation}
\label{Beq-RD1} \ddot b + \frac{3}{2t}\dot b +
\left(-\frac{1}{2(m_B t)^2}+\frac{1}{\sqrt{1+b^2}}\right)b = 0.
\end{equation}
As in the case of $B$ domination, the amplitude of $b$ shortly
diminishes and becomes smaller than unity.   Then the equation
(\ref{Beq-RD1}) is approximated as a linear equation in $b$ to
yield the usual damped-oscillation;
\begin{equation}
b(t) \approx \sqrt{\frac{2}{\pi}}\frac{1}{(m_B t)^{3/4}} \left[
C_1 \cos\left( m_B t +\frac{3\pi}{8}\right) + C_2 \sin\left( m_B t
+\frac{3\pi}{8}\right) \right]
\end{equation}
for $m_B t \gg 1$.  In the regime, the energy density of the
$B$-matter is found to be
\begin{equation}
 \rho \propto \frac{1}{(m_B t)^{3/2}} \propto 1/a^3 .
\end{equation}
That is, for $m_B t\gg1$, the oscillation dominates over the
expansion and the oscillating $B$ field behaves like the usual
matter  during the radiation dominated era.  As a consequence the
radiation will be overtaken by the $B$-matter eventually and the
cosmological evolution follows then the result of the subsection
3.2.

\setcounter{equation}{0}
\section{\large\bf Conclusion}

We have investigated the cosmology with the antisymmetric tensor
field in a D-brane universe.  A peculiar feature of this system is
that the antisymmetric tensor field becomes massive due to its
coupling to the massless U(1) gauge field on the brane and thus
its cosmological consequences are drastically different from those
of the massless antisymmetric tensor field.  Analyzing the Bianchi
type I cosmology describing the simplest form of an anisotropic
universe, we found time-dependent homogeneous solutions exhibiting
the matter-like (pressureless) behavior of the massive
antisymmetric tensor field ($B$-matter). For instance, the
$B$-matter amplitude oscillating with the frequency of $m_B$ gets
damped due to expansion. The energy density scales like $\rho
\propto  1/a^3$ and the averaged pressure becomes vanishingly
small. As a consequence, the initial anisotropy dies away rapidly
and the isotropic universe is recovered at late time.

\section*{Acknowledgements}
This work was supported by the grant KRF-2002-070-C00022 (E.J.C.),
No.~R01-2004-000-10520-0 from the Basic Research Program of the
Korea Science \& Engineering Foundation (H.B.K), and Astrophysical
Research Center for the Structure and Evolution of the Cosmos
(ARCSEC) (Y.K.). H.B.K and Y.K thank KIAS, and Y.K. also thanks
to E-Ken, Nagoya University for the hospitality
during the visit for which a part of this work was done.

\end{document}